\documentclass[aps,prd,superscriptaddress,twocolumn]{revtex4-2}
\usepackage{enumerate}

\usepackage{amsmath}
\usepackage{amssymb}
\usepackage{amsfonts}
\usepackage{amscd}
\usepackage{mathrsfs}

\usepackage{latexsym}
\usepackage{bm}

\usepackage{graphicx}
\usepackage{epsfig}
\usepackage{extarrows}
\usepackage{hhline,multirow}
\usepackage{dcolumn}
\usepackage{url}
\usepackage[colorlinks=true,linkcolor=blue]{hyperref}
\usepackage{color}
\usepackage{indentfirst}
\usepackage{subfigure}
\usepackage{psfrag}
\usepackage{slashed}
\usepackage{float}
\usepackage{setspace}
\usepackage{multirow}

\begin{document}
\title{Solution of lepton $g-2$ anomalies with nonlocal QED}
\author{Hang Li}
\affiliation{Institute of High Energy Physics, CAS, P. O. Box
	918(4), Beijing 100049, China}
\affiliation{School of Physical Sciences, University of Chinese Academy of Sciences, Beijing 101408, China}

\author{P. Wang}
\affiliation{Institute of High Energy Physics, CAS, P. O. Box
	918(4), Beijing 100049, China}
\affiliation{School of Physical Sciences, University of Chinese Academy of Sciences, Beijing 101408, China}

\begin{abstract}
The explanation for lepton $g-2$ anomalies is provided with the nonlocal QED which is the simple extension of the standard model. This solution is based on the same gauge symmetry as QED without introducing any new particles and interactions. The correlation functions in the nonlocal strength tension and lepton-photon interaction make it possible to explain the discrepancies of both $\Delta a_\mu$ and $\Delta a_e$. With the same approach, the discrepancy of anomalous magnetic moment of $\tau$ lepton from standard model is estimated in the range $1.19\times10^{-5}$ to $3.21\times10^{-3}$ which is covered by the current experimental uncertainty.
\end{abstract}

\maketitle

\section*{Introduction}

The anomalous magnetic moments of electron and muon, $a_e$ and $a_\mu$ are the most precisely determined quantities in particle physics. Theoretically, $a_e$ and $a_\mu$ are calculated within the standard model (SM) which contain several contributions: quantum electrodynamics (QED), electroweak (EW), hadronic vacuum polarization (HVP) and hadronic light-by-light (HLbL) \cite{Aoyama3}. The QED and EW parts are known very well perturbatively. For example, the QED contribution to the anomalous magnetic moments of electron and muon is known up to five-loop order \cite{Volkov,Aoyama,Aoyama2}. The others, being hadronic terms, are less well known and are estimated using various techniques, including data from the experiments \cite{Keshavarzi} and lattice calculations \cite{Blum,Chakraborty,Blum2,Borsanyi}. The updated theoretical predictions of $a_e$ and $a_\mu$
from standard model so far are $a_e^{\text{SM}}=1159652182.032(720)\times10^{-12}$ \cite{Aoyama} and $a_\mu^{\text{SM}}=116591810(43)\times10^{-11}$ \cite{Aoyama3}, respectively.

The recent measurement of the muon anomalous magnetic moment by the E989 experiment at Fermilab shows 
\begin{equation}
\Delta a_\mu^{\text{FNAL}} = a_\mu^{\text{FNAL}} - a_\mu^{\text{SM}} = (230 \pm 69) \times 10^{-11},
\end{equation}
which is a $3.3 \sigma$ discrepancy from the SM prediction \cite{Abi}. Combined with the previous E821 experiment at BNL \cite{Bennett}, the result leads to a $4.2 \sigma$ discrepancy \cite{Abi}
\begin{equation}
\Delta a_\mu = a_\mu^{\text{FNAL}+\text{BNL}} - a_\mu^{\text{SM}} = (251 \pm 59) \times 10^{-11}.
\end{equation}
For electron, the most accurate measurement of $a_e$ has been carried out by the Harvard group and the discrepancy from SM was $2.4 \sigma$ \cite{Hanneke}
\begin{equation}\label{eq:e1}
\Delta a_e^{\text{B}}=a_e^{\text{exp}}-a_e^{\text{SM,B}}=(-87\pm36)\times10^{-14}.
\end{equation}
However, a new determination of the fine structure constant $\alpha$ \cite{Morel}, obtained from the measurement at Laboratoire Kastler Brossel (LKB) with $^{87}{\text {Rb}}$, improves the accuracy by a factor of $2.5$ compared to the previous best measurement with $^{137}{\text{Cs}}$ atoms at Berkeley \cite{Parker}. With the new $\alpha$, the SM prediction for the electron magnetic moment is $1.6 \sigma$ lower than the experimental data, i.e.,
\begin{equation}\label{eq:e2}
\Delta a_e^{\text{LKB}}=a_e^{\text{exp}}-a_e^{\text{SM,LKB}}=(48\pm30)\times10^{-14}.
\end{equation}
In Eqs.~(\ref{eq:e1}) and (\ref{eq:e2}), the superscripts B and LKB are written referring to the different results due to different $\alpha$. 
Interestingly, two measurements give similar magnitude of the discrepancy, but with opposite signs. The small difference of $\alpha$ does not affect $\Delta a_\mu$ because it is much larger than $\Delta a_e$.
As standard model predictions almost match perfectly all other experimental information, the deviation in one of the most precisely measured quantities in particle physics remains a mystery and inspires the imagination of model builders. Certainly, the accurate determination of $\alpha$ will be important to constraint the theoretical solutions for the lepton $g-2$ anomalies \cite{Andreev,Keung}. 

Compared with the nucleon case where anomalous magnetic moments can be well described by effective field theory with the magnetic term explicitly included in the Lagrangian, the magnetic interaction between lepton and photon is incompatible with QED due to its non-renormalizability. Therefore, theoretical solutions to the discrepancy of lepton magnetic moments were proposed without exception by introducing new particles, symmetries and interactions beyond standard model \cite{Saez,Bai,Borah,Li,Wang,Chakraborti,Aboubrahim,Cox,Dey,Ghorbani,Perez,Ban,Arcadi,Jana,Cadeddu}. 
Most of the theoretical explanations focus on the muon anomalous magnetic moment and 
it is somewhat challenging to find a common beyond-standard-model origin to resolve both the muon and the electron $g-2$ anomalies, because of their large magnitude difference and possible opposite signs. There have been some new physical models which attempt to explain the muon and electron $g-2$ simultaneously \cite{Davoudiasl,Crivellin,Liu,Endo,Badziak,Calibbi,Chen,Dutta,Botella,Dorsner,Chun,Li2,Han}.
Since the new introduced particles have nearly no visible effects to other physical observations, they are usually related to the candidates for dark matter. 

In this work, we will provide a new view to possible solutions of the lepton $g-2$ discrepancy. The unique advantage of our approach is that the Lagrangian has the same gauge symmetry and same interaction as QED, except it is nonlocal. The $g-2$ discrepancies of electron and muon from the experiments can be explained simultaneously without introducing any new particles. The nonlocal QED is inspired by the nonlocal effective field theory, which has been successfully applied to study nucleon structure \cite{He,Salamu1,Salamu2,Yang}. The nonlocal characteristic could be the general property for all the interactions and therefore, it is straightforward to extend the SM to nonlocal QED to study lepton ``structure". The key characteristic of nonlocal theory is that on the one hand, the correlation functions in nonlocal Lagrangian make the loop integral ultraviolet convergent. On the other hand, it has the same local gauge symmetry as the corresponding local theory. With the improvement of experimental and theoretical accuracy in particle physics, the nonlocal feature could become more and more important.

\section*{Anomalous Magnetic Moment}

We start from a ``minimum" extension of the standard model. The nonlocal QED Lagrangian for studying lepton anomalous magnetic moments can be written as
\begin{eqnarray}
\mathcal{L}_{\text{QED}}^{\text{nl}}&=&\int d^4 a \Big[\bar{\psi}(x)(i\slashed{D}-m)\psi(x) \nonumber \\
&& -\frac14 F^{\mu\nu}(x)F_{\mu\nu}(x+a)F_\gamma(a) \Big],
\end{eqnarray}
where the covariant derivative $D_\mu = \partial_\mu + ie A_\mu(x+a)F_l(a)$.
In the above Lagrangian, $F_l(a)$ and $F_\gamma(a)$ are the correlation functions. 
If they are chosen to be $\delta$ functions, the nonlocal Lagrangian will change back to the local one.
The above Lagrangian is invariant under the following U(1) gauge transformation
\begin{equation} \label{eq:gauge-trans}
\psi(x)\to e^{i\alpha(x)}\psi(x), ~~ A_\mu(x)\to A_\mu(x)-\frac{1}{e}\partial_\mu \alpha'(x),
\end{equation}
where 
\begin{equation}
\alpha(x)=\int d^4 a\alpha'(x+a)F_l(a). 
\end{equation}
Different from our previous work, where $F_l(a)$ was introduced in calculating the Pauli form factors of leptons \cite{He2}, here, we introduce another correlation function $F_\gamma(a)$ to the strength tensor of photon field, which does not break the local gauge symmetry since $F_{\mu\nu}(x)$ and $F_{\mu\nu}(x+a)$ are both invariant under the transformation of Eq.~(\ref{eq:gauge-trans}). 
The correlation functions $F_\gamma(a)$ and $F_l(a)$ have opposite effects on the loop integrals and it is crucial to explain electron and muon anomalous magnetic moments simultaneously.
The function $F_\gamma(a)$ results in the modified photon propagator expressed as 
\begin{equation}
D_{\mu\nu} (k) = \frac{-i g_{\mu\nu}}{(k^2+i\epsilon)\tilde{F}_\gamma(k)}, 
\end{equation}
while $F_l(a)$ generate the momentum dependent vertex $ie\gamma^\mu\tilde{F}_l(k)$,
where $\tilde{F}_\gamma(k)$ and $\tilde{F}_l(k)$ are the Fourier transformations of $F_\gamma(a)$ and $F_l(a)$, respectively.

The renormalized lepton charge is defined as the Dirac form factor with zero momentum transfer $F_1(0)$, which is 1 due to the relationship $F_1^{\text{loop}}(0)=-\frac{\Sigma(\slashed{p})}{d\slashed{p}}$ \cite{He2},
where
\begin{eqnarray}
F_1^{\text{loop}}(0) &=& -2ie^2 \int \frac{d^4k}{(2\pi)^4} 
\frac{\tilde{F}_l^2(k)}{\tilde{F}_\gamma(k)} \nonumber \\
&\times& 
\frac{2m_l^4+m_l^2k^2 - 2m_l^2k \cdot p-2(k\cdot p)^2}{k^2(k^2 - 2k \cdot p)^2m_l^2} 
\end{eqnarray}
and
\begin{equation}
\Sigma(\slashed{p})=-ie^2 \int \frac{d^4k}{(2\pi)^4}\gamma_\mu\frac{1}{\slashed{p}-\slashed{k}-m_l}\gamma^\mu\frac{1}{k^2} \frac{\tilde{F}_l^2(k)}{\tilde{F}_\gamma(k)}.
\end{equation}
Compared with the local QED, the identity between the vertex and self-energy is not changed because the introduced correlators are lepton-momentum independent. This is also guaranteed by the local gauge symmetry of the nonlocal Lagrangian.
The anomalous magnetic moment $a_l$ is defined as the Pauli form factor with zero momentum transfer $F_2(0)$ expressed as
\begin{align}
a_l&=-2ie^2 \int \frac{d^4k}{(2\pi)^4}\frac{\tilde{F}_l^2(k)}{\tilde{F}_\gamma(k)}\left\{ \frac{(k \cdot q)^2}{q^2k^2(k^2-2k \cdot p)^2} \right.\notag\\
& \left.\left. -\frac{(k^2 + 2k \cdot p)m_l^2-3(k \cdot p)^2}{k^2(k^2 - 2k \cdot p)^2m_l^2}\right\}\right|_{q^2=0}.
\end{align}

In our previous calculations, $\tilde{F}_{l}(k)$ was chosen to be dipole form as
\begin{eqnarray}
\tilde{F}_l(k)&=&\left(\frac{\Lambda_l^2-m_l^2}{\Lambda_l^2-k^2}\right)^2.
\end{eqnarray}
Similarly, for the other regulator $\tilde{F}_\gamma(k)$ in the photon propagator, we choose the form to be 
\begin{eqnarray}\label{eq:fgama}
\tilde{F}_\gamma(k)&=&\left(\frac{\Lambda_\gamma^2}{\Lambda_\gamma^2-k^2}\right)^n.
\end{eqnarray}
In the above regulators, $\Lambda_l$ and $\Lambda_\gamma$ are the free parameters. In the numerical calculations, we choose $n$ to be 2, 3 and 4 to see whether both the electron and muon anomalous magnetic moments can be explained. The regulator $\tilde{F}_l(k)$ makes the loop integral more convergent, while $\tilde{F}_\gamma(k)$ makes the loop integral more divergent. Since there are two $\tilde{F}_l(k)$ and one $\tilde{F}_\gamma(k)$, the loop integral is more convergent for $n=2$ and 3 than that in SM. It will have the same convergent behavior as SM if $n=4$. We will not try the case for $n>4$ because it will make the loop integral more divergent. 

After the momentum integral and Taylor expansion, $a_l^{\text{2,n}}$ can be obtained, where the superscript $(2,n)$ means $\tilde{F}_l(k)$ is a dipole function and the power of $\tilde{F}_\gamma(k)$ in Eq.~(\ref{eq:fgama}) is $n$.
For the dipole, tripole and tetrapole correlators of $\tilde{F}_\gamma(k)$, $a_l^{\text{2,2}}$, $a_l^{\text{2,3}}$ and $a_l^{\text{2,4}}$ are expressed as
\begin{eqnarray}\label{eq:di}
a_l^{\text{2,2}} &=& \frac{\alpha}{2\pi}\left[1+\frac43\left(\frac{1}{\Lambda_\gamma^2}-\frac{2}{\Lambda_l^2}\right)m_l^2\right. \nonumber \\
&+& \left. 4 \left(\frac{1}{\Lambda_\gamma^4}-\frac{8}{\Lambda_\gamma^2\Lambda_l^2}+\frac{10}{\Lambda_l^4}\right)m_l^4\log\frac{\Lambda_l}{m_l}+{\cal O}(m_l^6)\right],
 \nonumber \\ 
a_l^{\text{2,3}} &=& \frac{\alpha}{2\pi}\left[1+\frac23\left(\frac{3}{\Lambda_\gamma^2}-\frac{4}{\Lambda_l^2}\right)m_l^2\right.
 \nonumber \\
&+& \left. \frac13\left(\frac{2\Lambda_l^2}{\Lambda_\gamma^6}-\frac{141}{2\Lambda_\gamma^4}+\frac{228}{\Lambda_\gamma^2\Lambda_l^2}-\frac{172}{\Lambda_l^4}\right)m_l^4 \right.
 \nonumber \\
&+& \left. 4\left(\frac{3}{\Lambda_\gamma^4}-\frac{12}{\Lambda_\gamma^2\Lambda_l^2}+\frac{10}{\Lambda_l^4}\right)m_l^4\log\frac{\Lambda_l}{m_l} +{\cal O}(m_l^6)\right], \nonumber \\
a_l^{\text{2,4}}&=&\frac{\alpha}{2\pi}\left[1+\frac83\left(\frac{1}{\Lambda_\gamma^2}-\frac{1}{\Lambda_l^2}\right)m_l^2\right. \nonumber \\
&+&\left.\frac13\left(\frac{\Lambda_l^4}{\Lambda_\gamma^8}+\frac{8\Lambda_l^2}{\Lambda_\gamma^6}-\frac{141}{\Lambda_\gamma^4}+\frac{304}{\Lambda_\gamma^2\Lambda_l^2}-\frac{172}{\Lambda_l^4}\right)m_l^4\right. \nonumber \\
&+&\left. 8\left(\frac{3}{\Lambda_\gamma^4}-\frac{8}{\Lambda_\gamma^2\Lambda_l^2}+\frac{5}{\Lambda_l^4}\right)m_l^4\log\frac{\Lambda_l}{m_l}+{\cal O}(m_l^6)\right], \nonumber \\
\end{eqnarray}
respectively.

From Eq.~(\ref{eq:di}), one can see that on the one hand, with the proper choices of the parameters $\Lambda_\gamma$ and $\Lambda_l$, the positive $\Delta a_\mu^{\text{nl}}$ and negative or positive $\Delta a_e^{\text{nl}}$ can be obtained, where $\Delta a_l^{\text{nl}}$ is the difference between nonlocal QED and standard model. On the other hand, at leading order, the discrepancy from the SM is proportional to the square of lepton mass $m_l$. It is very tempting to speculate about a simultaneous new-physics origin of the results with the same mechanism to explain the experimental discrepancies for both muon and electron. As pointed in Refs.~\cite{Calibbi,Botella}, although a common explanation has been shown to be possible, the model building task has proved non-trivial due to the scale $\Delta a_\mu/\Delta a_e \sim m_\mu^2/m_e^2$. Here with the nonlocal QED, the $g-2$ anomalies of electron and muon could be naturally explained. The specific parameters will be determined in the next section. Certainly, when $\Lambda_\gamma \rightarrow \infty$ and $\Lambda_l \rightarrow \infty$, $a_l$ will change back to the local result $\frac{\alpha}{2\pi}$.

\begin{figure}[hbt]
\centering
\includegraphics{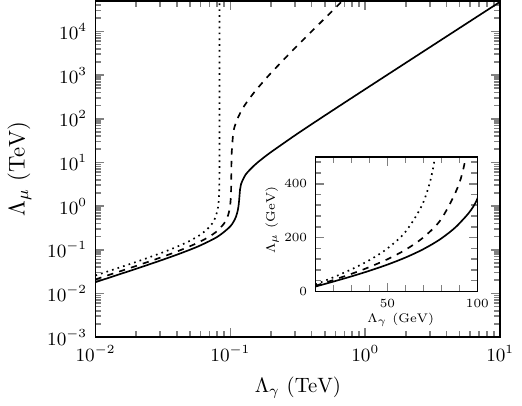}
\caption{The cutoff parameter $\Lambda_\mu$ versus $\Lambda_\gamma$. The dotted, dashed and solid lines are for dipole, tripole and tetrapole regulators in the photon propagator, respectively. The small figure at the corner is for the results at small $\Lambda_\gamma$.}
\label{fig:Lambda_mu-Lambda_g}
\end{figure}

\section*{Numerical results}

In this section, we will present the numerical results. First we determine the cutoff parameters $\Lambda$s with the experimental $\Delta a_e$ and $\Delta a_\mu$. 
We should mention that nonlocal QED itself can not pre-determine the form of the correlation functions and the parameters $\Lambda_\gamma$ and $\Lambda_l$. They reflect the properties of the particles and should be determined by the experiments. In practice, one can try and compare different correlators. The parameters $\Lambda$s will be determined to reproduce the experiment data. In this case, we determine $\Lambda_e$ and $\Lambda_\mu$ for a given $\Lambda_\gamma$ with the experimental $\Delta a_e$ and $\Delta a_\mu$ for the chosen correlators. 

In Fig.~\ref{fig:Lambda_mu-Lambda_g}, the obtained $\Lambda_\mu$ versus $\Lambda_\gamma$ are plotted. The dotted, dashed and solid lines are for dipole, tripole and tetrapole regulators in the photon propagator, respectively.
For a given $\Lambda_\gamma$, $\Lambda_\mu$ is obtained to get the experimental discrepancy. From the figure, one can see that $\Lambda_\mu$ increases with the increasing $\Lambda_\gamma$. When $\Lambda_\gamma$ is small, say dozens of GeV, the curves for (2,2), (2,3) and (2,4) cases are close to each other. For this range of small $\Lambda_\gamma$, the $m_\mu^2$ terms in Eq.~(\ref{eq:di}) are dominant for $\Delta a_\mu^{\text{nl}}$. The obtained $\Lambda_\mu$ are of the same order of magnitude. With the increase of $\Lambda_\gamma$, a transition region around 100 GeV appears, where the dominant contribution changes from $m_\mu^2$ terms to $m_\mu^4$ terms. 

For the (2,2) case, when $\Lambda_\gamma$ is larger than about 80 GeV, the $m_\mu^2$ term will be much smaller than the experimental $\Delta a_\mu$. The $m_\mu^4 \log\frac{\Lambda_\mu}{m_\mu}$ term will contribution the most part of $\Delta a_\mu^{\text{nl}}$. Because $m_\mu$ is very small compared with the cutoff parameters $\Lambda$s and the dominant contribution is proportional to $\log \frac{\Lambda_\mu}{m_\mu}$, the obtained $\Lambda_\mu$ is very large. When $\Lambda_\mu$ is infinite, the magnetic moment $a_\mu$ is $\log$ divergent. In other words, for any large $\Lambda_\gamma$, there always exist a $\Lambda_\mu$ to make $\Delta a_\mu^{\text{nl}}$ same as experimental data, though $\Lambda_\mu$ could be extremely large. 

For the (2,3) and (2,4) cases, the situations are similar while the $m_\mu^4$ terms are dominant at a little larger $\Lambda_\gamma$. Besides the $m_\mu^4\log\frac{\Lambda_\mu}{m_\mu}$ term,
there are two other terms proportional to $\frac{\Lambda_\mu^2}{\Lambda_\gamma^6}m_\mu^4$ and $\frac{\Lambda_\mu^4}{\Lambda_\gamma^8}m_\mu^4$ for (2,3) and (2,4) cases, respectively. They have much more important contributions to the discrepancy of magnetic moment than the $m_\mu^4\log\frac{\Lambda_\mu}{m_\mu}$ term when $\Lambda_\gamma$ is larger than about 100 GeV. 
As a result, for a fixed large $\Lambda_\gamma$ the obtained $\Lambda_\mu$ for (2,4) case is smaller than that for (2,3) case and both of them are much smaller than that for (2,2) case. For example, for the (2,4) case, the obtained $\Lambda_\mu$ are 
17.2 TeV, 117.7 TeV and 475.9 TeV for $\Lambda_\gamma$ 200 GeV, 500 GeV and 1 TeV, respectively.

\begin{figure}[hbt]
\centering
\includegraphics{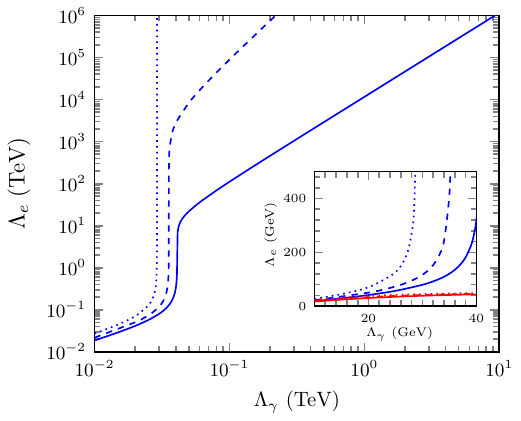}\label{fig:Lambda_e-Lambda_g}
\caption{The cutoff parameter $\Lambda_e$ versus $\Lambda_\gamma$. The dotted, dashed and solid lines are for dipole, tripole and tetrapole regulators in the photon propagator, respectively. The small figure at the corner is for the results at small $\Lambda_\gamma$. The red and blue lines are for negative $\Delta a_e^{\text{B}}$ and positive $\Delta a_e^{\text{LKB}}$, respectively.}
\label{fig:Lambda_e-Lambda_g}
\end{figure}

Same as for the muon, for the electron, the obtained $\Lambda_e$ versus $\Lambda_\gamma$ is plotted in Fig.~\ref{fig:Lambda_e-Lambda_g}. The red and blue lines are the results for experimental $\Delta a_e^{\text{B}}$ and $\Delta a_e^{\text{LKB}}$, respectively. For the negative 
$\Delta a_e^{\text{B}}$, the obtained $\Lambda_e$ are very small which are around 40 GeV for all the (2,2), (2,3) and (2,4) cases. $\Lambda_e$ is not sensitive to $\Lambda_\gamma$. In this case, the dominant contribution to $\Delta a_e^{\text{nl}}$ is from $-\frac{m_e^2}{\Lambda_e^2}$ term. The negative terms with the order of $m_e^4$ are highly suppressed due to the fact that $\Lambda_e$ and $\Lambda_\gamma$ are both on the denominator. This is different from the positive $\Delta a_\mu$ case, where the positive $m_\mu^4$ terms could have large contribution as long as $\Lambda_\mu$ is large enough for any fixed $\Lambda_\gamma$.

For the positive $\Delta a_e^{\text{LKB}}$, the situation is comparable with that for the muon case. 
When $\Lambda_\gamma$ is small, the $m_e^2$ terms are dominant. 
With the increase of  $\Lambda_\gamma$, the dominant contribution will change from $m_e^2$ terms to $m_e^4$ terms when $\Lambda_\gamma$ is larger than about 30 GeV. For the (2,2) case, again, the $m_e^4\log\frac{\Lambda_e}{m_e}$ term makes the obtained $\Lambda_e$ very large. For (2,3) and (2,4) cases, the obtained $\Lambda_e$
are much smaller than that for (2,2) case. Compared with the muon case, the obtained $\Lambda_e$ at the same $\Lambda_\gamma$ are much larger than $\Lambda_\mu$ for all the (2,2), (2,3) and (2,4) regulators. This is because the experimental value $\Delta a_\mu/\Delta a_e^{\text{LBK}}\sim m_\mu^2/m_e^2$. If $\Lambda_\mu = \Lambda_e$, we will have $\Delta a_\mu^{\text{nl}}/\Delta a_e^{\text{nl}}\sim m_\mu^4/m_e^4$. To get the experimental $\Delta a_e^{\text{LBK}}$, $\Lambda_e$ should be larger than $\Lambda_\mu$. In other words,
though the dominant contribution for $\Delta a_e^{\text{nl}}$ is from the $m_e^4$ term, the larger $\Lambda_e$ than $\Lambda_\mu$ makes $\Delta a_\mu^{\text{nl}}/\Delta a_e^{\text{nl}}\sim m_\mu^2/m_e^2$.

Therefore, for the electron, the different experimental data lead to different cutoff parameter $\Lambda_e$.
For a given $\Lambda_\gamma$, with the increasing $\Lambda_e$ from $\Lambda_e^{\text{B}}$ to $\Lambda_e^{\text{LKB}}$, $\Delta a_e^{\text{nl}}$ will increase continuously from negative $\Delta a_e^{\text{B}}$ to positive $\Delta a_e^{\text{LKB}}$. The accurate determination of the fine structure constant $\alpha$ is necessary to fix $\Lambda_e$ unambiguously.

From the above discussions, it is clear the $g-2$ discrepancies of muon and electron can be explained in the nonlocal QED with the  correlators. For small $\Lambda_\gamma$, say around dozens of GeV, the $g-2$ discrepancies $\Delta a_\mu$, $\Delta a_e^{\text{B}}$ and $\Delta a_e^{\text{LKB}}$ can all be reproduced with the corresponding $\Lambda_l$ which is of the similar magnitude
as $\Lambda_\gamma$. However, the introduced correlators in nonlocal QED should not change the conclusions for other high energy processes, since this is the only confirmed discrepancy from SM. Therefore, $\Lambda_\gamma$ and $\Lambda_l$ should be large enough to get same results as SM within the experimental uncertainty for other observables. For large $\Lambda_\gamma$, say larger than hundreds GeV or 1 TeV, both $\Delta a_\mu$, and $\Delta a_e^{\text{LKB}}$ can be reproduced with large $\Lambda_l$. However, we can not get the negative value $\Delta a_e^{\text{B}}$ with large $\Lambda_e$. Though $\Delta a_e^{\text{B}}$ is not possible to be reproduced with large $\Lambda_\gamma$ and $\Lambda_e$, we can still get negative value of $\Delta a_e^{\text{nl}}$ with smaller magnitude than $\Delta a_e^{\text{B}}$. For example, $\Delta a_e^{\text{nl}}$ is $-1.82\times10^{-15}$ with $\Lambda_\gamma=\Lambda_e= 1$ TeV.

\begin{figure}[hbt]
\centering
\includegraphics{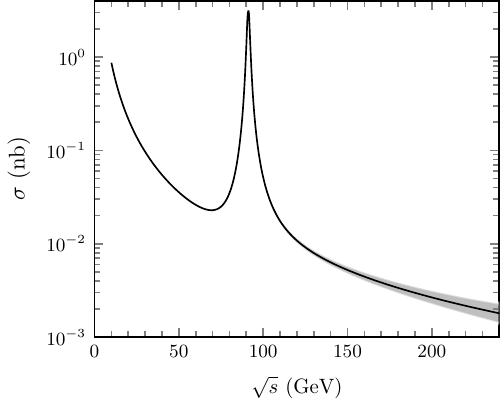}
\caption{The $\sqrt{s}$ dependence of the cross section of $e^+e^-\to\mu^+\mu^-$ at leading tree level with $\gamma$ and $Z$ exchanges. The solid line is the SM result. The narrow band is the result of nonlocal QED with $\Lambda_\gamma=1$ TeV and $\Lambda_l\geq 1$ TeV.}
\label{fig:ee-mumu}
\end{figure}

To see the effect of the nonlocal QED on the other processes with large $\Lambda$s, in Fig.~\ref{fig:ee-mumu}, the cross section of $e^+ e^- \rightarrow \mu^+ \mu^-$ versus $\sqrt{s}$ is plotted as an example. The solid line is for the leading tree level result of SM with $\gamma$ and $Z$ exchanges, while the narrow band is for the nonlocal result with $\Lambda_\gamma=1$ TeV
and $\Lambda_l \geq 1$ TeV. The propagator of $Z$ boson is modified as that of photon with the same correlator. From the figure, one can see, when the collision energy is not large, the difference is not visible. The pole at the mass of $Z$ boson does not change. A slight difference appears when the energy is larger than about 120 GeV. The difference increases with the increasing $\sqrt{s}$. Therefore, we can conclude if the chosen $\Lambda$s are large enough, say larger than hundreds GeV or 1 TeV, the nonlocal results are consistent with the experimental data as well as the SM.  

\begin{figure}[ht]
\centering
\includegraphics{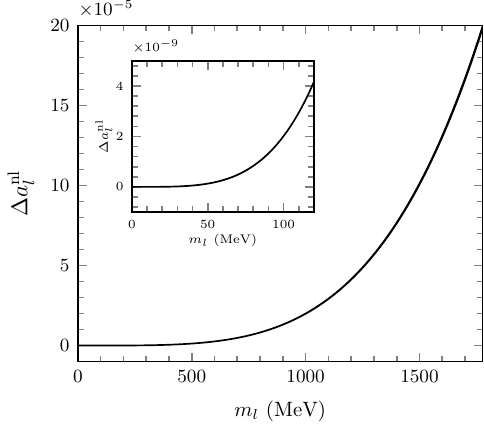}
\caption{The calculated discrepancy of lepton anomalous magnetic moment $\Delta a_l^{\text{nl}}$ versus lepton mass $m_l$ with $\Lambda_\gamma=1$ TeV and $\Lambda_\mu=\Lambda_\gamma$. The small figure at the corner is for the result at small lepton mass.}
\label{fig:magvsm}
\end{figure}

We have shown for sufficiently large $\Lambda$s, the nonlocal QED can explain the $g-2$ anomaly of muon and the recent result of the positive discrepancy of electron anomalous magnetic moment $\Delta a_e^{\text{LKB}}$. The discrepancy $\Delta a_e^{\text{nl}}$ can also be negative, but with smaller magnitude than $\Delta a_e^{\text{B}}$. In addition, our nonlocal QED have little affect on the other processes as long as $\Lambda$s are chosen to be large enough.

Now we discuss the lepton mass dependence of $\Delta a_l^{\text{nl}}$. In Fig.~\ref{fig:magvsm}, the discrepancy $\Delta a_l^{\text{nl}}$ versus lepton mass $m_l$ is plotted for correlators with $\Lambda_\gamma=1$ TeV. A small figure is plotted to show the result at small lepton mass. Because $\Lambda_e$ is not well determined due to two different measurements, $\Lambda_l$ is fixed to be $\Lambda_\mu$ obtained by the experimental $\Delta a_\mu$ with $\Lambda_\gamma=1$ TeV. 
There is no visible difference among the three regulators for
$n=2$, 3, and 4 in Eq.~(\ref{eq:fgama}).
When $m_l=m_\mu$, $\Delta a_l^{\text{nl}}$ equals to the experimental discrepancy $\Delta a_\mu$, while $\Delta a_l^{\text{nl}}$ is between $\Delta a_e^{\text{LKB}}$ and $\Delta a_e^{\text{B}}$ when $m_l=m_e$.
The discrepancy increases with the increasing lepton mass and it is $1.98\times 10^{-4}$ when $m_l$ is at the mass of $\tau$ lepton. 

\begin{figure}[ht]
\centering
\includegraphics{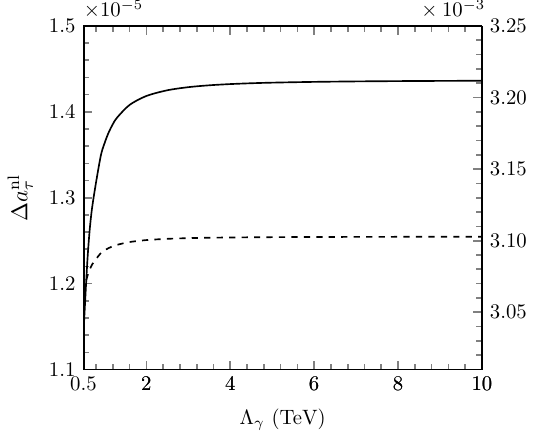}\caption{The calculated discrepancy of anomalous magnetic moment $\Delta a_\tau^{\text{nl}}$ versus $\Lambda_\gamma$. The solid line is for the upper limit of $\Delta a_\tau^{\text{nl}}$ with $\Lambda_\tau = 2 \Lambda_\mu$ (right axis), while the dashed line is for the lower limit with $\Lambda_\tau = \frac{\Lambda_\mu}{2}$ (left axis).}
\label{fig:magtauvslg}
\end{figure}

Based on the nonlocal explanation for the $g-2$ anomalies of muon and electron, we will give an estimation for $\Delta a_\tau^{\text{nl}}$ using the same approach. 
For a given $\Lambda_\gamma$, $\Lambda_\mu$ can be well determined by the experimental data. 
For $\tau$ lepton, we vary $\Lambda_\gamma$ from 500 GeV to 10 TeV and assume $\Lambda_\tau$ is in the range $\frac{\Lambda_\mu}{2}$ to $2 \Lambda_\mu$. In Fig.~\ref{fig:magtauvslg}, the calculated $\Delta a_\tau^{\text{nl}}$ versus $\Lambda_\gamma$ is plotted. 
The solid and dashed lines are for the upper and lower limits of $\Delta a_\tau^{\text{nl}}$ with $\Lambda_\tau = 2 \Lambda_\mu$ and $\Lambda_\tau = \frac{\Lambda_\mu}{2}$, respectively.
The difference among the three regulators is also included when we estimate the range of $\Delta a_\tau^{\text{nl}}$. 
For this range of parameters, the dominant contribution to the discrepancy of magnetic moment of muon and tau is from the $m_l^4$ terms. As a result, $\Delta a_\tau^{\text{nl}} / \Delta a_\mu^{\text{nl}} \sim m_\tau^4/m_\mu^4$.
$\Delta a_\tau^{\text{nl}}$ increases with the increasing $\Lambda_\gamma$ and it is not sensitive to $\Lambda_\gamma$ when $\Lambda_\gamma$ is larger than about 2 TeV. The estimated range of $\Delta a_\tau^{\text{nl}}$ is from $1.19 \times 10^{-5}$ to $3.21\times 10^{-3}$. 

\section*{Summary}

In summary, we studied the anomalous magnetic moments of leptons in nonlocal QED which is the simple extension of the standard model. The nonlocal part of the strength tension of QED generates the modified photon propagator, while the nonlocal electron-photon interaction leads to the momentum dependent vertex. With large enough cutoff parameters $\Lambda_\gamma$ and $\Lambda_l$ in the correlators, the experimental discrepancies $\Delta a_\mu$ and $\Delta a_e^{\text{LKB}}$ can be well obtained. Though the discrepancy $\Delta a_e^{\text{B}}$ can not be reproduced, the calculated $\Delta a_e^{\text{nl}}$ can also be negative, but with smaller magnitude than $\Delta a_e^{\text{B}}$.
Since the introduced $\Lambda$s are sufficiently large, the nonlocal effects to the other physical quantities are negligible.
Therefore, we provide an explanation for the $g-2$ anomalies of muon and electron without introducing any new particles and symmetries. Certainly, the accurate determination of the fine structure constant in the future is very important to constraint the theoretical solutions of $g-2$ anomalies. With the same approach, $\Delta a_\tau^{\text{nl}}$ is estimated to be in the range $1.19\times10^{-5}$ to $3.21\times10^{-3}$, which is obviously covered by the large experimental uncertainty $(-0.052,0.013)$. This unique solution to lepton $g-2$ anomalies is quite different from all the other methods. Nonlocal approach in fundamental interactions could become a complementary direction for precisely testing standard model.

\section*{Acknowledgments}

This work is supported by the NSFC under Grant No.~11975241.


\begin{thebibliography}{81}
\bibitem{Aoyama3} T.~Aoyama $et$ $al$., Phys. Rept. {\bf 887}, 1 (2020).
\bibitem{Volkov} S.~Volkov, Phys.~Rev.~D {\bf 100}, 096004 (2019).
\bibitem{Aoyama} T.~Aoyama, $et$ $al$., Phys.~Rev.~D {\bf 97}, 036001 (2018).
\bibitem{Aoyama2} T.~Aoyama,$et$ $al$., Phys.~Rev.~Lett. {\bf 109}, 111807, (2012).
\bibitem{Keshavarzi} A.~Keshavarzi, D.~Nomura, and T.~Teubner, Phys.~Rev.~D {\bf 97}, 114025 (2018). 
\bibitem{Blum} T.~Blum $et$ $al$., (RBC and UKQCD Collaborations), Phys.~Rev.~Lett. {\bf 121}, 022003 (2018)
\bibitem{Chakraborty} B.~Chakraborty, $et$ $al$., (Fermilab Lattice, HPQCD, and MILC Collaborations), Phys.~Rev.~D {\bf 98}, 094503 (2018).
\bibitem{Blum2} T.~Blum $et$ $al$., Phys.~Rev.~Lett. {\bf 124}, 132002 (2020).
\bibitem{Borsanyi} Sz.~Borsanyi $et$ $al$., Nature 593, 7857, 51 (2021).
\bibitem{Abi} B~.Abi {\em et al}. (Muon $g-2$ Collaboration), Phys.~Rev.~Lett. {\bf 126}, 141801 (2021).
\bibitem{Bennett} G.~W.~Bennett {\em et al}. (Muon $g-2$ Collaboration), Phys.~Rev.~D {\bf 73}, 072003 (2006).
\bibitem{Hanneke} D.~Hanneke {\em et al}., Phys.~Rev.~Lett. {\bf 100}, 120801 (2008).
\bibitem{Morel} L.~Morel, Z.~Yao, P.~Clade, and S.~Guellati-Khelifa, Nature {\bf 588}, 61 (2020).
\bibitem{Parker} R.~H.~Parker, C.~Yu, W.~Zhong, B.~Estey and H.~Muller, Science {\bf 360}, 191 (2018).
\bibitem{Andreev} Y. M. Andreev $et$ $al$., Phys.~Rev.~Lett. {\bf 126}, 211802 (2021).
\bibitem{Keung} W. Y. Keung, D. Marfatia, and P. Y. Tseng,
LHEP {\bf 2021}, 209 (2021).
\bibitem{Saez} B.~D.~Saez and K.~Ghorbani, Phys.~Lett.~B {\bf 823}, 136750 (2021).
\bibitem{Bai} Y.~Bai, S.~J.~Lee, M.~Son, and F.~Ye, arXiv:2106.15626. 
\bibitem{Borah} D.~Borah, A.~Dasgupta, and D.~Mahanta, Phys.~Rev.~D {\bf 104}, 075006 (2021).
\bibitem{Li} Z.~Li, G.~L.~Liu, F.~Wang, J.~M.~Yang, and Y.~Zhang, arXiv:2106.04466.
\bibitem{Wang} F.~Wang, L.~Wu, Y.~Xiao, J.~M.~Yang, and Y.~Zhang, Nucl.~Phys.~B {\bf 970}, 115486 (2021)
\bibitem{Chakraborti} M.~Chakraborti, L.~Roszkowski, and S.~Trojanowski, JHEP {\bf 05}, 252 (2021).
\bibitem{Aboubrahim} A.~Aboubrahim, P.~Nath, and R.~M.~Syed, JHEP {\bf 06}, 002 (2021).
\bibitem{Cox} P.~Cox, C.~Han, and T.~T.~Yanagida, Phys.~Rev.~D {\bf 98}, 055015 (2018).
\bibitem{Dey} A.~Dey, J.~Lahiri, and B.~Mukhopadhyaya, arXiv:2106.01449.
\bibitem{Ghorbani} K.~Ghorbani, arXiv:2104.13810.
\bibitem{Perez} P.~F.~Perez, C.~Murgui, and A.~D.~Plascencia, Phys.~Rev.~D {\bf 104} 035041 (2021).
\bibitem{Ban} K.~Ban $et$ $al$., arXiv:2104.06656.
\bibitem{Arcadi} G.~Arcadi, A.~S.~De Jesus, T.~B.~De Melo, F.~S.~Queiroz, and Y.~S.~Villamizar, arXiv:2104.04456.
\bibitem{Jana} S.~Jana, P.~K.~Vishnu, W.~Rodejohann, and S.~Saad, Phys.~Rev.~D {\bf 102}, 075003 (2020).
\bibitem{Cadeddu} M.~Cadeddu, N.~Cargioli, F.~Dordei, C.~Giunti, and E.~Picciau, Phys.~Rev.~D {\bf 104}, 011701 (2021).
\bibitem{Davoudiasl} H.~Davoudiasl and W.~J.~Marciano,~Phys.~Rev.~D {\bf 98}, 075011 (2018).\bibitem{Crivellin} A.~Crivellin, M.~Hoferichter, and P.~Schmidt-Wellenburg, Phys.~Rev.~D {\bf 98}, 113002 (2018).
\bibitem{Liu} J.~Liu, C.~E.~M.~Wagner, and X.~P.~Wang, JHEP {\bf 03}, 008 (2019).
\bibitem{Endo} M.~Endo and W.~Yin, JHEP {\bf 08}, 122 (2019).
\bibitem{Badziak} M.~Badziak and K.~Sakurai, JHEP {\bf 10}, 024 (2019).
\bibitem{Calibbi} L.~Calibbi, M.~L.~Lopez-Ibanez, A.~Melis, and O.~Vives, JHEP {\bf 06}, 087 (2020).
\bibitem{Chen} C.~H.~Chen and T.~Nomura, Nucl.Phys.B {\bf 964}, 115314 (2021).
\bibitem{Dutta} B.~Dutta, S.~Ghosh, and T.~Li, Phys.~Rev.~D {\bf 102}, 055017 (2020).
\bibitem{Botella} F.~J.~Botella, F.~Cornet-Gomez, and M.~Nebot, Phys.~Rev.~D {\bf 102}, 035023 (2020).
\bibitem{Dorsner} I.~Dorsner, S.~Fajfer, and S.~Saad, Phys.~Rev.~D {\bf 102} 075007 (2020).
\bibitem{Chun} E.~J.~Chun and T.~Mondal, JHEP {\bf 11}, 077 (2020).
\bibitem{Li2} S.~P.~Li, X.~Q.~Li, Y.~Y.~Li, Y.~D~.Yang, and X.~Zhang, JHEP {\bf 01}, 034 (2021).
\bibitem{Han} X.~F.~Han, T.~J.~Li, H.~X.~Wang, L.~Wang, and Y.~Zhang, arXiv:2104.03227.
\bibitem{He} F.~He and P.~Wang, Phys.~Rev.~D {\bf 97}, 036007 (2018).
\bibitem{Salamu1} Y.~Salamu, C.~R.~Ji, W.~Melnitchouk, A.~W.~Thomas, and P.~Wang, Phys.~Rev.~D {\bf 99}, 014041 (2019).
\bibitem{Salamu2} Y.~Salamu, C.~R.~Ji, W.~Melnitchouk, A.~W.~Thomas, P.~Wang, and X.~G.~Wang, Phys.~Rev.~D {\bf 100}, 094026 (2019).
\bibitem{Yang} M.~Y.~Yang and P.~Wang, Phys.~Rev.~D {\bf 102}, 056024 (2020).
\bibitem{He2} F.~He and P.~Wang, ~Eur.~Phys.~J.~Plus {\bf 135}, 156 (2020).

\end{thebibliography}
\end{document}